\documentclass[3p,times,twocolumn]{elsarticle}

\usepackage{ecrc}

\volume{00}

\firstpage{1}

\journalname{Nuclear Physics B Proceedings Supplement}

\runauth{}

\jid{nuphbp}

\jnltitlelogo{Nuclear Physics B Proceedings Supplement}

\usepackage{amssymb}

\usepackage[figuresright]{rotating}
\usepackage{epsfig}
\usepackage{graphics}
\usepackage[usenames,dvipsnames,svgnames,table]{xcolor}

\def\lsim{\mathrel {\vcenter {\baselineskip 0pt \kern 0pt
    \hbox{$<$} \kern 0pt \hbox{$\sim$} }}}
\def\gsim{\mathrel {\vcenter {\baselineskip 0pt \kern 0pt
    \hbox{$>$} \kern 0pt \hbox{$\sim$} }}}

\begin{document}

\begin{frontmatter}



\dochead{}

\title{Higgs production constraints on anomalous fermion couplings}


\author[label1]{Alper Hayreter}
\author[label2]{ and German Valencia}

\address[label1]{Department of Natural and Mathematical Sciences, Ozyegin University, 34794 Istanbul Turkey.}

\address[label2]{Department of Physics, Iowa State University, Ames, IA 50011.}

\begin{abstract}
Certain anomalous fermion-gauge boson couplings, such as the flavor diagonal anomalous color magnetic (CMDM) and color electric (CEDM) dipole moments of quarks are not fully gauge invariant under the SM. Restoring gauge invariance with an elementary Higgs doublet implies that they also contribute to Higgs boson production at the LHC and we study the corresponding constraints that can be placed on them. In a similar manner we study the constraints that can be placed on the $\tau$-lepton anomalous magnetic moment, electric dipole moment, weak dipole moments, and dimension eight gluonic couplings at the LHC.
\end{abstract}


\end{frontmatter}


\section{Introduction}
\label{}
\begin{figure}[thb]
\includegraphics[width=0.5 \textwidth]{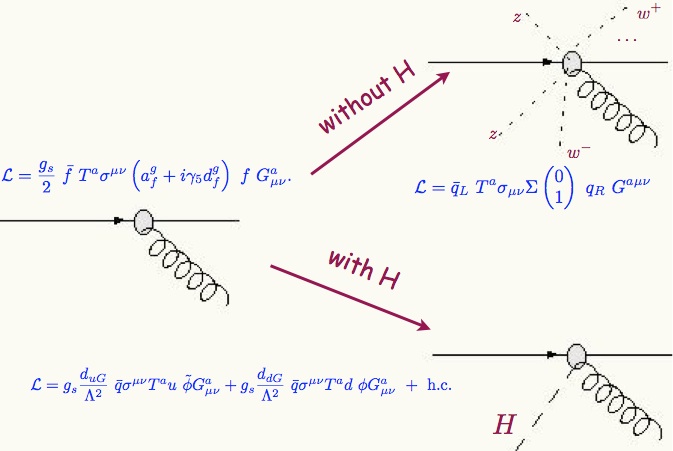}
\caption{The dim 5 operator can be made gauge invariant without a Higgs in which case it is related to vertices with additional $w^\pm,z$, or with a Higgs and then it is related to vertices that include a Higgs boson.}
\label{gaugeinv}
\end{figure}

New physics couplings frequently discussed are the dipole couplings of fermions. For quarks, one has couplings to gluons referred to as anomalous color magnetic (CMDM) and color electric (CEDM) dipole moments of the form,
\begin{eqnarray}
{\cal L}=\frac{g_s}{2}\ d_{qG}\ \bar{f}_L\ T^a\sigma^{\mu\nu} \ f_R\ G^a_{\mu\nu} + {\rm h.~c.}
\label{defcoup}
\end{eqnarray}
As it stands, however, this operator is not fully gauge invariant under the standard model (SM). There are a few ways to think about this:
\begin{itemize}
\item We only wrote the unitary gauge version of a Lagrangian in which the spontaneously broken gauge symmetry is nonlinearly realized. This is appropriate if the observed 126~GeV scalar is not a fundamental Higgs \cite{Callan:1969sn,Peccei:1990uv,Buchalla:2013rka}.
\item We need to fix the gauge invariance with a scalar doublet with a vev as in the SM \cite{Buchmuller:1985jz,Grzadkowski:2010es}.
\end{itemize}
each of these two scenarios has different consequences for phenomenology as illustrated in Fig.~\ref{gaugeinv}. In this talk we examine the latter case following the results of our papers \cite{Hayreter:2013kba,Hayreter:2013vna}.

\section{quark anomalous couplings}

\begin{figure*}[htb]
\includegraphics[width=0.5 \textwidth]{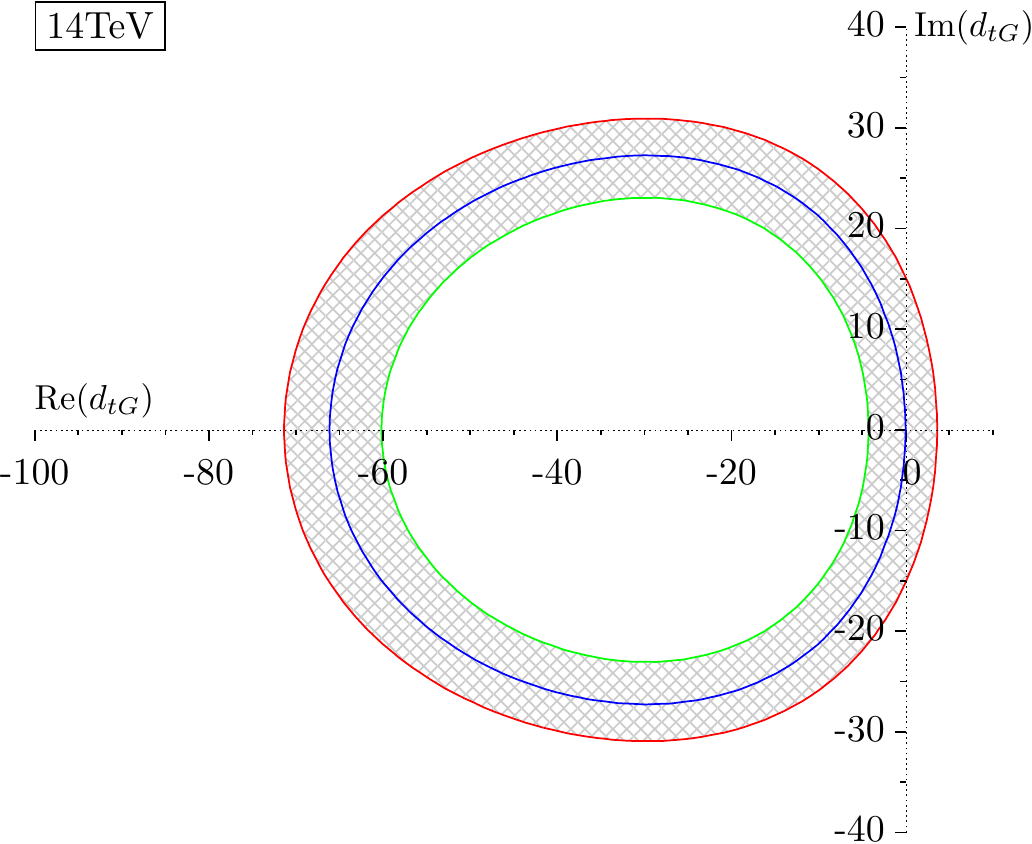}\includegraphics[width=0.5 \textwidth]{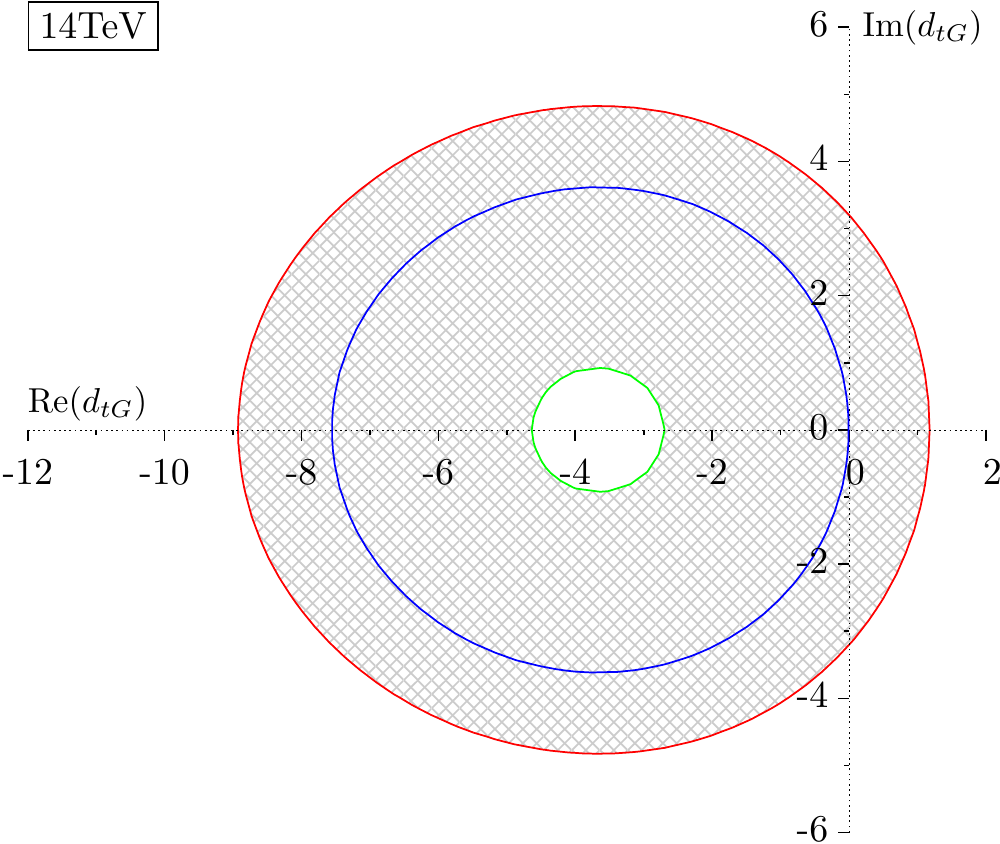}
\caption{$1\sigma$ bounds on CMDM and CEDM of the top-quark that could be obtained from measuring the cross-sections for $pp\to t\bar{t}$ (left) and $pp\to t\bar{t}h$ (right) at 14TeV at the LHC.}
\label{ttvstth}
\end{figure*}

For the remainder of this talk we will assume that the 126~GeV scalar is the Higgs boson, in which case the gauge invariant version of Eq.~\ref{defcoup} for the top-quark is, (with obvious notational extension to all other quarks)
\begin{eqnarray}
{\cal L} = g_s\frac{d_{tG}}{\Lambda^2}\ \bar{q}_{3L}\sigma^{\mu\nu}T^a t_R\  \tilde\phi G^a_{\mu\nu}   +\ {\rm h.c.}
\label{GIanocoup}
\end{eqnarray}
and the conventional anomalous couplings in terms of $d_{tG}$ are given by
\begin{eqnarray}
{\cal L}&=&\frac{g_s}{2}\ \bar{t}\ T^a\sigma^{\mu\nu}\left(a_t^g+i\gamma_5 d_t^g\right) \ t\ G^a_{\mu\nu}\nonumber \\
a_{t}^g &=& \frac{\sqrt{2} \ v}{\Lambda^2}{\rm ~Re}(d_{tG}),\, 
d_{t}^g = \frac{\sqrt{2} \ v}{\Lambda^2}{\rm ~Im}(d_{tG}).
\label{anomcouplings}
\end{eqnarray}
These couplings are subject to the usual constraints from top-quark pair production \cite{Atwood:1992vj,Cheung:1995nt,Choi:1997ie,Sjolin:2003ah,Martinez:2007qf,Antipin:2008zx, Hioki:2009hm,Larkoski:2010am,Englert:2012by,HIOKI:2011xx,Kamenik:2011wt,Gupta:2009wu,Baumgart:2012ay,Biswal:2012dr,Hioki:2013hva}, but they also affect and can be constrained by Higgs production associated with a top-quark pair \cite{DeRujula:1990db,Choudhury:2012np,Degrande:2012gr,Einhorn:2013tja,Englert:2014oea,Bramante:2014gda,Englert:2014uua,Zhang:2014rja,Ellis:2014dva,Masso:2014xra}. We compare the two types of constraints in Fig.~\ref{ttvstth} for the 14~TeV LHC. For this comparison we have implemented the couplings into {\tt MadGraph5} \cite{MadGraph} with the aid of {\tt FeynRules} \cite{Christensen:2008py} at LO. The resulting cross-sections are quartic polynomials in the anomalous couplings with only even powers of the CP-odd coupling. We simulate enough points to fit the polynomials that describe the new physics and its interference with the SM, and add them to the best known prediction for the SM as discussed in Ref.~\cite{Hayreter:2013kba}.
The bounds shown in Figure~\ref{ttvstth} arise from saturating the theoretical error of the NLO SM cross-sections. This assumes that the measurements will agree with the SM and that the theory error will dominate. Numerically it corresponds to a 14\% of $pp\to t\bar{t}$ \cite{Beneke:2011ys} and 15-18\% of $pp\to t\bar{t}h$ \cite{Dittmaier:2011ti} being attributed to new physics. We can see in the figure that $pp\to t\bar{t}h$ places better constraints for `natural' values of the couplings (values closer to 0); much better overall constraints (allowing destructive interference with the SM); and much better constraints for the CEDM (imaginary part). For constraints from 8 TeV data see Ref.~\cite{Hayreter:2013kba}.

\begin{figure*}[thb]
\includegraphics[width=0.5 \textwidth]{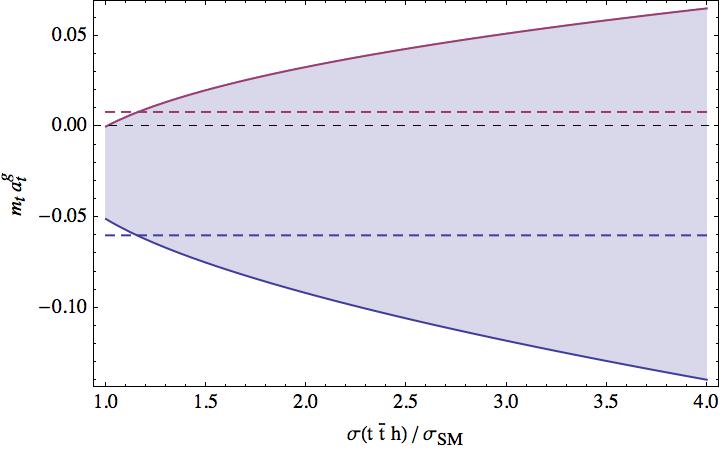}\includegraphics[width=0.5 \textwidth]{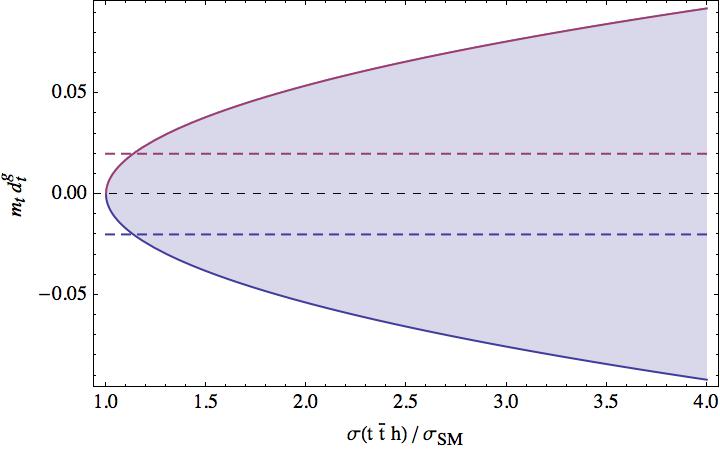}
\caption{Bounds on CMDM and CEDM of the top-quark that can be obtained from limiting the ratio $\sigma(pp\to t\bar{t}h)/SM$ at 14TeV at the LHC.}
\label{tthlimit}
\end{figure*}

The measurement of $\sigma(pp\to t\bar{t}h)$ is difficult and it is interesting to see the constraints that can be placed based on a limit only \cite{Chatrchyan:2013yea}. This is illustrated in Fig.~\ref{tthlimit} where the dashed lines show the $+15\%$ contours from Fig.~\ref{ttvstth}. 

It is well known that measurements of asymmetries in $pp\to t\bar{t}$ can improve the limits on these couplings. For example the CEDM can be constrained at $5\sigma$ with 10${\rm ~fb}^{-1}$ to be less than $0.1/m_t$ through T-odd asymmetries\cite{Sjolin:2003ah,Antipin:2008zx,Gupta:2009wu} at LHC14. Optimizing spin-correlations the CEDM and CMDM can be constrained at the $0.05/m_t$ and $0.03/m_t$ respectively with 20${\rm ~fb}^{-1}$ at LHC8\cite{Baumgart:2012ay}. In a similar manner, asymmetries in $pp\to t\bar{t}h$ can also do better than cross-section  measurements, at least in principle. In practice however, thousands of $\rm{fb}^{-1}$ would be needed to measure an asymmetry at the \% level.

\begin{figure}[thb]
\includegraphics[width=0.5 \textwidth]{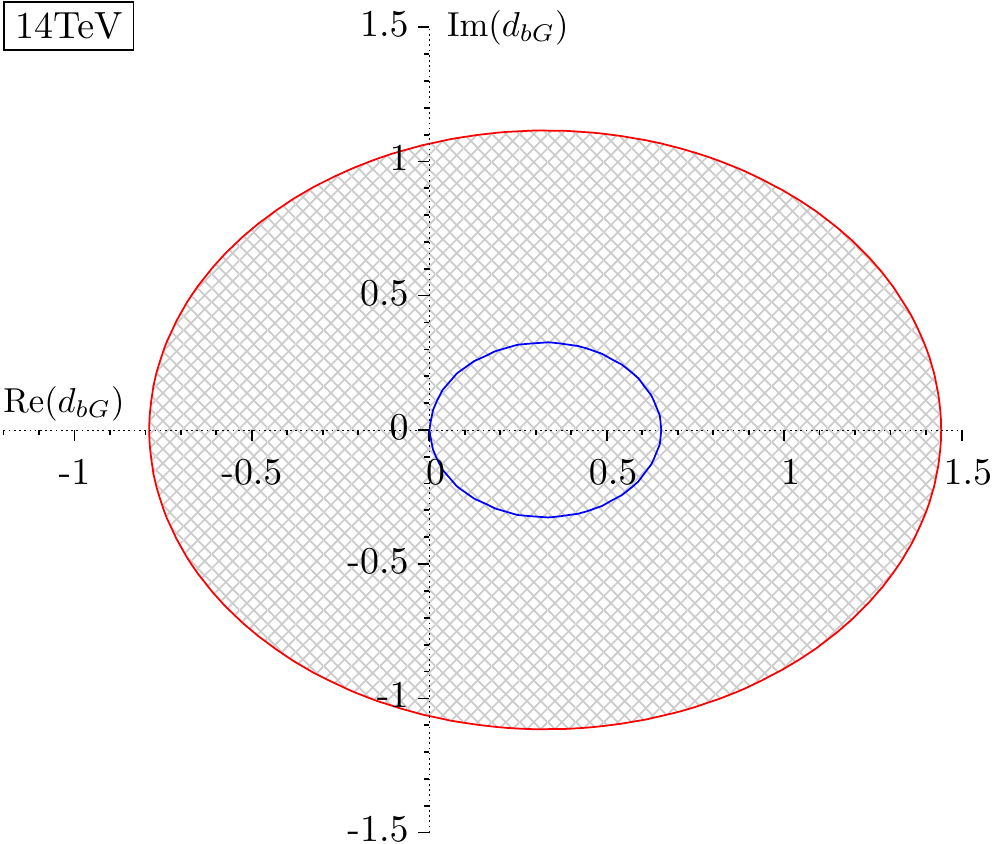}
\caption{Bounds on CMDM and CEDM of the bottom-quark that can be obtained from a 17\% error in $\sigma(pp\to b\bar{b}h)$ at 14TeV at the LHC.}
\label{bbh}
\end{figure}

For the case of the $b$-quark NP effects in $b\bar{b}$ production are overwhelmed by QCD and Higgs production in association with a $b\bar{b}$ pair becomes the best hope to constrain these couplings at the LHC. The process $\sigma(pp\to b \bar{b}hX)$ is being searched for in the context of non-SM Higgs, in particular for large $\tan\beta$ so that a bound could be obtained and compared to the SM NLO prediction. If a measurement is made and it agrees with the SM, the 17\% error in the theoretical prediction \cite{Dittmaier:2003ej} would yield the bounds shown in Fig.~\ref{bbh} \cite{Hayreter:2013kba}. 
For further study of this case see Ref.~\cite{Bramante:2014hua}.

\begin{figure}[thb]
\includegraphics[width=0.5 \textwidth]{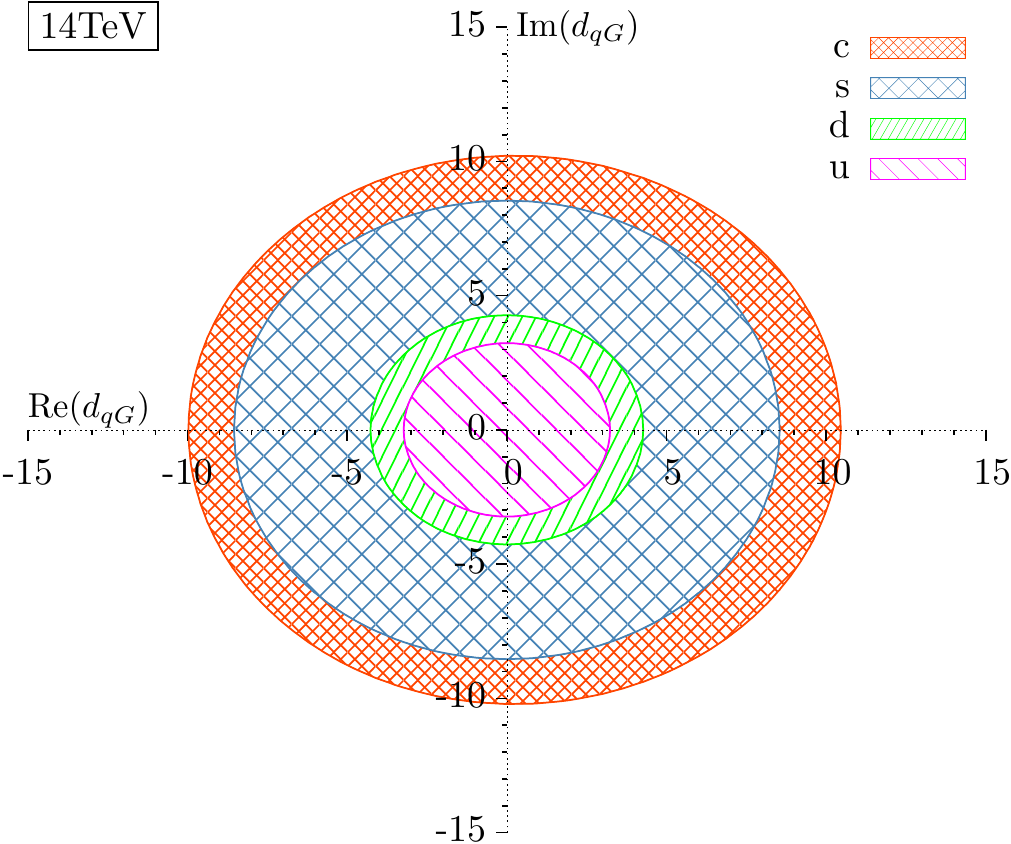}
\caption{Bounds on CMDM and CEDM of the  the light quarks from the $1\sigma$ error in $\sigma(pp\to hX)$ at 14TeV at the LHC.}
\label{qqh}
\end{figure}

For the light quarks, including charm, NP is again buried in QCD background except perhaps in processes with a Higgs. Here we perform the simplest analysis and constrain the anomalous couplings of the light quarks with the $1\sigma$ error in the  $\sigma(pp\to hX)$ at 14TeV at the LHC. We show the results in Fig.~\ref{qqh}. Within the SM the subprocesses $qg\to qh$ and $q\bar{q}\to hg$ are dominated by charm. In all cases the interference between the SM and NP (being proportional to the fermion mass) is negligible. The cross-sections are quadratic in the anomalous couplings so we generate enough MC points for different values of the NP couplings  to fit a quadratic form. To obtain our results we require the NP contributions to fall below the theoretical uncertainty of the dominant gluon fusion SM process. 
 This simple picture fails beyond LO where heavy quark loops give larger SM contributions \cite{Field:2003yy,Campbell:2012am}. Our bounds could be improved by using the Higgs plus one jet mode which is more sensitive to new physics at higher $p_T$ but this analysis has not been carried out yet.

In Table~\ref{t:results} we summarize our results and compare them to indirect constraints from the neutron edm. We also translate the constraints into an effective new physics scale that the LHC can reach at the $1\sigma$ sensitivity, in all cases between 1 and 3 TeV. For the indirect constraints for $u$, $d$, $s$ we use the neutron (or $\Lambda$) edm and the quark model. For $c,b,t$ the constraints arise from the Weinberg three gluon operator as in Ref.~\cite{DeRujula:1990db} but notice that a more recent estimate for the top-quark is larger than the value listed on the table by an order of magnitude \cite{Kamenik:2011dk}. There is also a recent analysis for the charm case \cite{Sala:2013osa}.
\begin{table*}[htb]
\caption{Summary of results for 1$\sigma$ bounds that can be placed on the CEDM at LHC and indirect constraints from neutron or $\Lambda$ edms.}
\begin{center}
\begin{tabular}{|c|c|c|c|c|}
\hline
\color{red} LHC Process & \color{red}CMDM & \color{red}CEDM & \color{red}$\Lambda$ (TeV) & \color{red} $n^\gamma_d$ \\ \hline
$\sigma(t\bar{t})$ 8 ~TeV& $-0.034\lsim m_ta_t^g\lsim 0.031$ &  $|m_t d_t^g|\lsim 0.12$ & (1.5, .7)& $2.4\times10^{-4}$ \\  \hline
$\sigma(t\bar{t})$ 14 ~TeV& $-0.029\lsim m_ta_t^g\lsim 0.024$ &  $|m_t d_t^g|\lsim 0.1$ & (1.5, .7)&\\  \hline
$A_1(t\bar{t})$ 14 TeV&- &  $|m_t d_t^g|\lsim 0.009$ &(-, 2.5) & \\  \hline
$\sigma(t\bar{t}h)$ 14~TeV& $-0.016\lsim m_ta_t^g\lsim 0.008$ &  $|m_t d_t^g|\lsim 0.02$ & (2, 1.7) &\\  \hline
$A_{1,2}(t\bar{t}h)$ 14~TeV& - &  $|m_t d_t^g|\lsim 0.007$ &(-, 3) & \\  \hline
$\sigma(b\bar{b}h)$ 14~TeV& $-1.3 \times 10^{-4}\lsim m_ba_b^g\lsim 2.4\times 10^{-4}$ &  $|m_b d_b^g|\lsim 1.7 \times 10^{-4}$& 2.7 & $2\times10^{-8}$ \\  \hline
$\sigma( hX)$ 8~TeV& $|a_u^g |\lsim 3.5 \times 10^{-4}\ {\rm GeV}^{-1}$ &  $|d_u^g |\lsim 3.5 \times 10^{-4}\ {\rm GeV}^{-1}$ & 1 & $1.8\times10^{-11}\ {\rm GeV}^{-1}$\\  \hline
$\sigma(hX)$ 14~TeV& $|a_u^g |\lsim 1.2 \times 10^{-4}\ {\rm GeV}^{-1}$ &  $|d_u^g |\lsim 1.2 \times 10^{-4}\ {\rm GeV}^{-1}$ &1.7 & \\  \hline
$\sigma(hX)$ 14~TeV& $|a_d^g |\lsim 1.6 \times 10^{-4}\ {\rm GeV}^{-1}$ &  $|d_d^g |\lsim 1.6 \times 10^{-4}\ {\rm GeV}^{-1}$ &1.5 &$1.8\times10^{-11}\ {\rm GeV}^{-1}$ \\  \hline
$\sigma( hX)$ 14~TeV& $|a_s^g |\lsim 3.3 \times 10^{-4}\ {\rm GeV}^{-1}$ &  $|d_s^g |\lsim 3.3 \times 10^{-4}\ {\rm GeV}^{-1}$ &1  &$0.1\ {\rm GeV}^{-1}$ $\Lambda^\gamma_d$ \\  \hline
$\sigma(hX)$ 14~TeV& $|a_c^g |\lsim 3.9 \times 10^{-4}\ {\rm GeV}^{-1}$ &  $|d_c^g |\lsim 3.9 \times 10^{-4}\ {\rm GeV}^{-1}$ &1&$4.7\times10^{-10}\ {\rm GeV}^{-1}$ \\  \hline
\end{tabular}
\end{center}
\label{t:results}
\end{table*}

\section{$\tau$-lepton anomalous couplings}

A similar study can be carried out for leptons. We first consider the dipole-type couplings of the leptons \cite{Barr:1988mc}
\begin{eqnarray}
{\cal L}&=&\frac{e}{2}\ \bar{\ell}\ \sigma^{\mu\nu}\left(a_\ell^\gamma+i\gamma_5 d_\ell^\gamma \right) \ \ell \ F_{\mu\nu}\nonumber \\
&+&\frac{g}{2\cos\theta_W}\ \bar{\ell}\ \sigma^{\mu\nu}\left(a_\ell^Z+i\gamma_5 d_\ell^Z \right) \ \ell \ Z_{\mu\nu}.
\label{defzdm}
\end{eqnarray}
which gauge invariance turns into 
\begin{eqnarray}
{\cal L} = g\frac{d_{\ell W}}{\Lambda^2}\ \bar{\ell}\sigma^{\mu\nu}\tau^i e\  \phi W^i_{\mu\nu} + g^\prime\frac{d_{\ell B}}{\Lambda^2}\ \bar{\ell}\sigma^{\mu\nu}e  \ \phi B_{\mu\nu} 
\label{ginvedm}
\end{eqnarray}
We also consider the enhanced dimension 8 operators
\begin{eqnarray}
{\cal L} = \frac{g_s^2}{\Lambda^4}\left(d_{\tau G} \ G^{A\mu\nu}G^A_{\mu\nu} \bar \ell_L \ell_R \phi  +d_{\tau \tilde{G}} \ G^{A\mu\nu} \tilde G^A_{\mu\nu} \bar \ell_L \ell_R \phi \right)
\label{taugluon}
\end{eqnarray}
where $G^A_{\mu\nu}$ is the gluon field strength tensor and $\tilde G^{A\mu\nu} = (1/2)\epsilon^{\mu\nu\alpha\beta}G^A_{\alpha\beta}$ its dual. These dimension 8 couplings are normally neglected as they are suppressed by two additional powers of the NP scale relative to the dimension 6 anomalous couplings. However, the effects of these particulars operators, `lepton-gluonic couplings'  that couple a lepton pair directly to gluons, are enhanced at the LHC due to the larger parton luminosities \cite{Potter:2012yv}.  Similar operators for the flavor violating case have also been discussed \cite{Petrov:2013vka}.

A possible observable to constrain these couplings is a deviation of the di-lepton cross-section from Drell-Yan in the large invariant mass region ($m_{\ell\ell}>120$~GeV) \cite{Chatrchyan:2012hd,Aad:2012gm}. It is also possible to study the $Z$ region \cite{Hayreter:2013vna}. Numerically we concentrate on $\tau$-leptons because existing constraints for muons and electrons are much stronger and many models exist where effects in the $\tau$-lepton are important \cite{delAguila:1990jg,Goozovat:1991nu,delAguila:1991rm,Cornet:1995pw,Vidal:1998jc,GonzalezSprinberg:2000mk,Bernabeu:2004ww,Bernabeu:2008ii,Bolanos:2013tda}. We assume that a measurement of the high invariant mass Drell-Yan cross-section for $\tau$ pairs at the 14\% level will be possible to obtain our constraints. The number 14\% is chosen because the current main systematic uncertainty in high invariant mass di-tau pairs at CMS ($m_{\tau\tau}>300$~GeV), arising from estimation of background, is in the range 6-14\% \cite{Chatrchyan:2012hd}. A second observable using the Higgs boson would be to constrain the cross-section $\sigma(pp\to \tau^+\tau^- h)$. Perhaps this could be done from searches for $pp\to Zh$ with a di-tau reconstruction of the Z.

In Table~\ref{tau:results} we summarize 1$\sigma$ constraints that can be placed on the $\tau$-lepton anomalous magnetic moment, electric dipole moment and weak dipole moments with a 14\% measurement of the Drell-Yan cross-section at LHC14. We compare them to the best existing constraints from Delphi \cite{Abdallah:2003xd}, Belle \cite{Inami:2002ah} and Aleph \cite{Heister:2002ik}. The results can be interpreted as a sensitivity to a NP scale $\Lambda \sim 0.5$~TeV. For comparison, the same measurement of the Drell-Yan cross-section constrains the NP scale of the dimension 8 gluonic couplings  $\Lambda \sim 1$~TeV \cite{Hayreter:2013vna}.
\begin{table*}[htb]
\caption{Summary of constraints for 1$\sigma$ bounds that can be placed on the $\tau$-lepton anomalous magnetic moment, electric dipole moment and weak dipole moments at LHC14 compared to existing bounds.}
\begin{center}
\begin{tabular}{|c|c|c|c|c|}
\hline
& $m_\tau a_\tau^V$ LHC-14 & $m_\tau a_\tau^V$ existing & $m_\tau d_\tau^V$ LHC-14 & $m_\tau d_\tau^V$ existing \\ \hline
$V=\gamma$ & (-0.0054,0.0060) & (-0.026,0.007) Delphi & (-0.0057,0.0057) &
(-0.002,0.0041) Belle \\ \hline
$V=Z$ & (-0.0018,0.0020) & (-0.0016,0.0016) Aleph & (-0.0017,0.0017) &
(-0.00067,0.00067) Aleph \\ \hline
\end{tabular}
\end{center}
\label{tau:results}
\end{table*}

We can now ask whether it is possible to improve on these LHC constraints using processes with a Higgs boson as was the case for the quarks.  Measuring the associated production $pp\to \tau^+\tau^-h$ will be very hard, so for now we ask instead what bounds on $\sigma(pp\to \tau^+\tau^-h)$ would be necessary in order to compete with a 14\% measurement of the Drell-Yan cross-section. We find that to improve the bounds on $d_\tau^{\gamma,Z}$ one would need to constrain $\sigma(pp\to \tau^+\tau^-h)\lsim 5$~fb for $m_{\tau\tau}>120$~GeV. In other words, one needs to constrain $\sigma(pp\to \tau^+\tau^-h)/\sigma_{SM} \lsim 50$. Similarly, for the gluonic couplings $d_{\tau G,\tilde{G}}$ one would need $\sigma(pp\to \tau^+\tau^-h)\lsim 50$~fb for $m_{\tau\tau}>120$~GeV or $\sigma(pp\to \tau^+\tau^-h)/\sigma_{SM} \lsim 500$.

\section{Summary}
 After the discovery of the Higgs boson there is a concerted effort to measure its couplings to other SM particles.
 We propose the use of processes with a Higgs to constrain anomalous couplings between SM fermions and gauge bosons.
With a fundamental 126 GeV Higgs breaking electroweak symmetry, gauge invariance relates these anomalous couplings to others between the same SM fermions and gauge bosons plus a Higgs. 
We have presented simple estimates for the constraints that can be expected at 14 TeV at the LHC. In some cases, the light quarks, it would not be possible to constrain the couplings at the LHC in processes without a Higgs.  In other cases (top-quark or $\tau$-lepton) the constraints  from processes with a Higgs are potentially much better.







\end{document}